\documentclass[preprint,superscriptaddress]{revtex4}
\usepackage[dvipdf]{graphicx}

\newcommand{\ket}[1] {\left| #1 \right\rangle}

\begin{document}

\title{Effect of disorder on transport properties in a tight-binding model for lead halide perovskites}

\author{S. Ashhab}
\affiliation{Qatar Environment and Energy Research Institute (QEERI), Hamad Bin Khalifa University (HBKU), Qatar Foundation, Doha, Qatar}
\author{O. Voznyy}
\affiliation{Department of Electrical and Computer Engineering, University of Toronto, Toronto, Ontario M5S 3G4, Canada}
\author{S. Hoogland}
\affiliation{Department of Electrical and Computer Engineering, University of Toronto, Toronto, Ontario M5S 3G4, Canada}
\author{E. H. Sargent}
\affiliation{Department of Electrical and Computer Engineering, University of Toronto, Toronto, Ontario M5S 3G4, Canada}
\author{M. E. Madjet}
\affiliation{Qatar Environment and Energy Research Institute (QEERI), Hamad Bin Khalifa University (HBKU), Qatar Foundation, Doha, Qatar}

\date{\today}


\begin{abstract}
The hybrid organic-inorganic lead halide perovskite materials have emerged as remarkable materials for photovoltaic applications. Their strengths include good electric transport properties in spite of the disorder inherent in them. Motivated by this observation, we analyze the effects of disorder on the energy eigenstates of a tight-binding model of these materials. In particular, we analyze the spatial extension of the energy eigenstates, which is quantified by the inverse participation ratio. This parameter exhibits a tendency, and possibly a phase transition, to localization as the on-site energy disorder strength is increased. However, we argue that the disorder in the lead halide perovskites corresponds to a point in the regime of highly delocalized states. Our results also suggest that the electronic states of mixed-halide materials tend to be more localized than those of pure materials, which suggests a weaker tendency to form extended bonding states in the mixed-halide materials and is therefore not favourable for halide mixing.
\end{abstract}

\maketitle

\section*{Introduction}

Hybrid organic-inorganic lead halide perovskites, with the representative material methylammonium lead iodide (CH$_3$NH$_3$PbI$_3$), have recently emerged as promising materials for optoelectronic applications, such as being the light-absorbing material in a solar cell \cite{Kojima,PerovskiteReview,PerovskiteTransportReview}. Among their attractive properties in this context are their near-optimal band gaps for solar-energy applications ($\sim 1.6$ eV), strong light absorption even at the band edge and good electronic transport properties. Another attractive property is that by substituting some of the iodine atoms by bromine atoms, the band gap can be increased to any desired value between 1.6 eV and 2.3 eV \cite{Edri,McMeekin}. This ability to obtain larger band gaps (especially around 1.9 eV) is highly desirable for tandem solar cell applications, where the optimal combination is that one active layer absorbs photons with energies above 1.9 eV while the other layer absorbs photons with energies above 1.1 eV.

\begin{figure}[h]
\includegraphics[width=8.0cm]{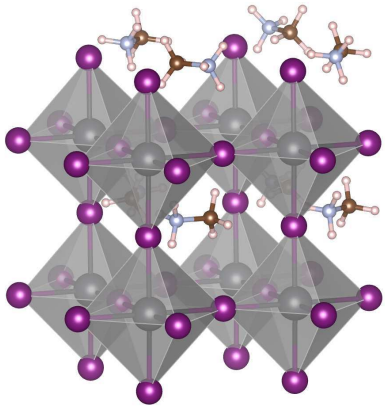}
\caption{Structure of the perovskite material MAPbI$_3$. The I atoms are located at the edges of the octahedra, with the Pb atoms at the centers of the octahedra and the MA molecules at the centers of the spaces between the octahedra. The disorder in the systems has two sources: (1) the random, and in fact temporally fluctuating, orientations of the MA molecules producing a random electrostatic potential landscape in the lattice and (2) the substitutional halide atoms, i.e.~Br atoms replacing some of the I atoms.}
\label{Fig:PerovskiteStructure}
\end{figure}

These materials have an almost inherent type of disorder. In the crystal structure of the materials, the CH$_3$NH$_3$ molecules (or MA molecules for short) reside in the spaces between the octahedra of the perovskite lattice, as shown in Fig.~\ref{Fig:PerovskiteStructure}. At room temperature these molecules are almost free to rotate and point in any direction, thus sampling all possible directions with almost uniform probability density \cite{Poglitsch,Mashiyama,Frost,Carignano2015,Leguy,Bakulin}. It is also believed that correlations between the orientations of the molecules at different sites are weak. In addition to a net charge, each MA cation possesses an electric dipole moment and contributes to the potential energy in its vicinity. As a result, the electrostatic potentials at the locations of the different metal or halide atoms are different from each other \cite{Ma}. Furthermore, the halide mixing that is used to adjust the band gap of the material leads to a disordered lattice, since it is believed that for example the iodine and bromine atoms are distributed randomly among the halide sites in the lattice.

There have been some discussions of the role of disorder in lead halide perovskites. For example, it has been experimentally shown that the disorder plays a role in the scattering of free carriers in these materials \cite{Laovorkiat}. Another experimental study investigated the nature of the dynamic disorder \cite{Yaffe}. There have also been studies on the fluctuations associated with the MA molecule orientation \cite{Carignano2015,Bakulin}. The disorder related to halide mixing was investigated in Ref.~\cite{Buin} in the context of identifying defect states in these materials.

Disorder is generally harmful to transport in quantum systems and can be expected to affect charge carrier diffusion, which occurs on the nanosecond timescale and involves the motion of carriers over thousands of unit cells. A dramatic manifestation of the adverse effects of disorder is Anderson localization \cite{Anderson}, where an increasing amount of disorder can lead to a phase transition from a metallic to an insulating state. This raises the question of why the mixed-halide perovskites have good transport properties in spite of being disordered. The answer to this question is unknown, and there are ongoing attempts to elucidate the origin of the unusual transport properties of these materials, as can be seen in Ref.~\cite{Berry}. In particular, it has been suggested that the nature of the atomic orbitals that form the valence and conduction bands is unfavourable for the formation of deep trap states where non-radiative recombination takes place \cite{Brandt}. In addition, a high mobility is important to extract carriers faster than they can recombine or be trapped, which is possible in a material where states remain largely delocalized despite the large degree of disorder caused by the random orientations of the organic molecules, compositional variations (iodide and bromide mixing, formamidinium [FA] and MA mixing, etc.), and structural imperfections (vacancies, etc.). Interestingly, Ref.~\cite{Ma} suggested that disorder could in fact be helping in electron-hole separation, which is an important step in the operation of a solar cell.

The effects of disorder on the electronic properties of these materials can be investigated using simulations of disordered structures. First-principles calculations with more than a few unit cells are in principle possible but computationally demanding \cite{Ma}. One can instead use a tight-binding model, which reduces the number of degrees of freedom per unit cell drastically and allows the simulation of larger systems \cite{Goringe,Agapito}. We take the model developed in Ref.~\cite{BoyerRichard}, including the parameters obtained there, as the starting point for our calculations. This simplification allows us to perform systematic simulations on systems containing 512 unit cells (effectively containing $\sim$6000 atoms) at a rather low computational cost.

\section*{Results}

\subsection*{Tight-binding model}

We consider a model where electrons can occupy and hop between the Pb and halide atomic sites, and at each site there are four orbitals available to it, one $s$ and three $p$ orbitals. More specifically, for MAPbI$_3$ the relevant orbitals are Pb 6$s$, Pb 6$p$, I 5$s$ and I 5$p$. We do not include any orbitals associated with the MA molecules, because these orbitals are far away from the valence band maximum (VBM) and conduction band minimum (CBM). We focus on the cubic phase, where the structure exhibits symmetry between the three axes x, y and z. In this case, the halide atoms are arranged to form perfect octahedra that are all aligned with each other, which keeps the number of system parameters (i.e. orbital energies and orbital overlaps) at the reasonable value 9. Spin-orbit coupling (SOC) is also included in the model, which adds two more parameters to the model, namely the SOC strengths for Pb and I. Treating the orbital energies and overlaps as fitting parameters, Boyer-Richard {\it et al.} obtained system parameters that reproduce the main features in the electronic band structure of MAPbI$_3$ \cite{BoyerRichard}. It is worth mentioning here that a recent study used a somewhat similar tight-binding model to investigate SOC effects in a different family of halide perovskite materials \cite{Kim}.

The model is described by the Hamiltonian:
\begin{equation}
\hat{H} = \sum_{i,\alpha} \epsilon_{i,\alpha} \hat{a}^{\dagger}_{i,\alpha} \hat{a}_{i,\alpha} - \sum_{i,j,\alpha,\beta} t_{i\alpha j\beta} \left( \hat{a}^{\dagger}_{i,\alpha} \hat{a}_{j,\beta} + h.c. \right),
\label{Eq:SpinChainHamiltonian}
\end{equation}
where $\epsilon_{i,\alpha}$ is the energy of orbital $\alpha$ (i.e.~$s$ or $p$) at site $i$, $t_{i\alpha j\beta}$ is the hopping strength between orbital $\alpha$ at site $i$ and orbital $\beta$ at site $j$, $\hat{a}_{i,\alpha}$ is the particle annihilation operator for orbital $\alpha$ at site $i$, and $\hat{a}^{\dagger}_{i,\alpha}$ is its hermitian conjugate. Spin-orbit coupling, which couples different $p$ orbitals and different spin states, for atom A (i.e.~Pb or I) is described by the Hamiltonian:
\begin{equation}
\hat{H}_{\rm SOC, A} = \frac{\lambda_A}{2} \left( \begin{array}{cccccc}
0 & 0 & -i & 0 & 0 & 1 \\
0 & 0 & 0 & i & -1 & 0 \\
i & 0 & 0 & 0 & 0 & -i \\
0 & -i & 0 & 0 & -i & 0 \\
0 & -1 & 0 & i & 0 & 0 \\
1 & 0 & i & 0 & 0 & 0
\end{array} \right),
\end{equation}
which we have written in the basis $\left\{\ket{p_x,\uparrow},\ket{p_x,\downarrow},\ket{p_y,\uparrow},\ket{p_y,\downarrow},\ket{p_z,\uparrow},\ket{p_z,\downarrow}\right\}$. In the absence of disorder, we set $\epsilon_{i,\alpha}$, $t_{i\alpha j\beta}$ and $\lambda_A$ to the values given in Ref.~\cite{BoyerRichard}.

Two types of disorder are present in the problem that we wish to study. The disorder in the orientations of the MA molecule can be simulated by introducing random fluctuations in the on-site energies $\epsilon_{i,\alpha}$, which we take to depend only on the location of the orbital and therefore to be independent of $\alpha$. A substitution of I by Br in general results in changes in the orbital energies as well as the hopping strengths between orbitals at neighboring sites \cite{MadjetThermalization}. We consider both of these mechanisms as possible causes of the changes that accompany halide substitution. Consequently, we explore two different ways to investigate the effect of Br substitution. In one set of calculations, we simulate mixed-halide materials by setting the Br orbital energies to lower values than the I orbitals, while keeping the hopping strengths fixed. We find that in order to obtain the correct band gap for MAPbBr$_3$ we must set the Br orbital energies 0.86 eV lower than those of I, and this is the value that we use in our simulations. In another set of of calculations, we fix the orbital energies and set the hopping strength between Pb and Br sites to be 0.73 of the value for Pb-I hopping, where as above the value 0.73 is chosen because it gives the correct band gap for MAPbBr$_3$.

An interesting question here relates to the definition of the ordered state. Since each MA molecule contributes a potential energy to the different orbitals depending on its orientation, the ordered state would have well-defined orientations of the molecules, leading to a nonzero potential felt by electrons at neighbouring atoms. Recent molecular dynamics simulations suggest some correlation between neighbouring MA molecules \cite{Frost,Carignano2015,Mattoni,Lahnsteiner,Meloni,Carignano2017}. The correlations are such that in one direction the MA molecules tend to be anti-parallel to each other and the other two directions neighbouring molecules tend to be perpendicular to each other. Such correlations imply that the MA molecules will not produce a macroscopic polarization that will lead to a large contribution to the electric field inside the crystal. In our model, we ignore any net contribution to the potential from the MA molecules in the ordered state and effectively treat the molecules as electron donors in that case.

Another question relates to the logic behind taking experimental parameters that are measured for a disordered system and using them to extract the parameters of the disorder-free model. The idea of this approach would be that there are two possible outcomes: if we find that the physical properties (and in particular the band structure) of the ordered and disordered states are only slightly different from each other, then the results justify our use of ignoring the difference between the ordered and disordered states when calculating the parameters of the model based on the properties of realistic, and hence disordered, systems. If on the other hand we find that the physical properties of the ordered and disordered states are drastically different, we would have to go back and reevaluate our basic assumptions. As we shall see below, one of our main conclusions is that realistic levels of disorder do not significantly modify the electronic states of the materials under study.

\subsection*{Inverse participation ratio}

Disorder typically has the effect of turning the extended energy eigenstates of perfectly ordered potentials into localized states. This localization in the energy eigenstates then translates into the transport properties, e.g.~resulting in a reduced mobility. A good quantity that can be used to analyze such localization properties is the inverse participation ratio (IPR), which for a given quantum state $\psi(i)$ is given by:
\begin{equation}
{\rm IPR} = \frac{\sum_{i=1}^NP(i)^2}{\left(\sum_{i=1}^NP(i)\right)^2}.
\end{equation}
where $P(i)$ is the occupation probability of site $i$, i.e.
\begin{equation}
P(i) = \left|\psi (i)\right|^2,
\end{equation}
and the index $i$ labels the different sites in the lattice and therefore runs over all the sites in the lattice \cite{Bell,Visscher,Ashhab,Magnetta}. Roughly speaking, the IPR for a state gives the inverse of the spatial extent (or volume) of the state. Each energy eigenstate has its IPR value, and as we shall discuss below, different parts of the band structure become localized at different levels of disorder.

The band structure that we obtain from our tight-binding model can be naturally divided into three parts. We shall analyze the mean values of the IPR for these three parts of the spectrum. The IPR for the filled states gives us an idea about the bonding strength in the material. In particular, having electrons in delocalized states can be interpreted as meaning that the electrons form bonds between the atoms, which indicates energetic stability of the material. We shall also analyze the IPR for the states at the VBM and CBM, which can serve as indicators for the transport properties (e.g.~the mobility) of holes and electrons. For the VBM and CBM calculations, we take the mean values for the states that lie within 0.1 eV from the band edges.

\subsection*{Computational details}

For disorder of strength $\Delta\epsilon$ in the on-site energies $\epsilon_{i,\alpha}$, we include a Gaussian-distributed random contribution to the orbital energies with the standard deviation of such an ensemble set to $\Delta\epsilon$. Our use of the Gaussian distribution is motivated by the central limit theorem. Each I atom is surrounded by four equally distant MA molecules in its immediate vicinity followed by larger numbers of additional MA molecules when we consider neighbouring unit cells. Each Pb atom has eight equally distant MA molecules serving as its nearest-neighbour molecules. These numbers are far from approaching infinity, but they are sufficiently large to suggest that regardless of the exact distribution of the potential energy fluctuation contributed by each molecule the sum of these contributions will have an overall shape resembling a Gaussian distribution. As mentioned above, we shift all four orbitals at a given site by the same amount. We apply the same rules for calculating the fluctuations to both Pb and I atoms. When simulating mixed-halide materials, we first generate configurations with the two halide species randomly distributed in the lattice (with a specified average concentrations $x$ and $1-x$) and then we either lower the orbital energies at the Br sites by 0.86 eV or reduce the coupling strengths between the Br and Pb sites by the factor 0.27, as explained above.

We investigate the properties of the system by considering a supercell composed of $8 \times 8 \times 8$ cubic unit cells. Since each cubic unit cell contains one Pb atom and three halide atoms and we are including four orbitals and two spin states for each atomic site, our supercell has 16,384 states that an electron can occupy. The size of the Hilbert space is therefore 16,384. Each basis state in the 16,384-dimensional Hilbert space corresponds to the electron being in a given orbital at a given atomic site in the supercell with spin state up or down. We set the quasi-momentum to zero in our calculations.

\subsection*{Calculation results}

\begin{figure}[h]
\includegraphics[width=15.0cm]{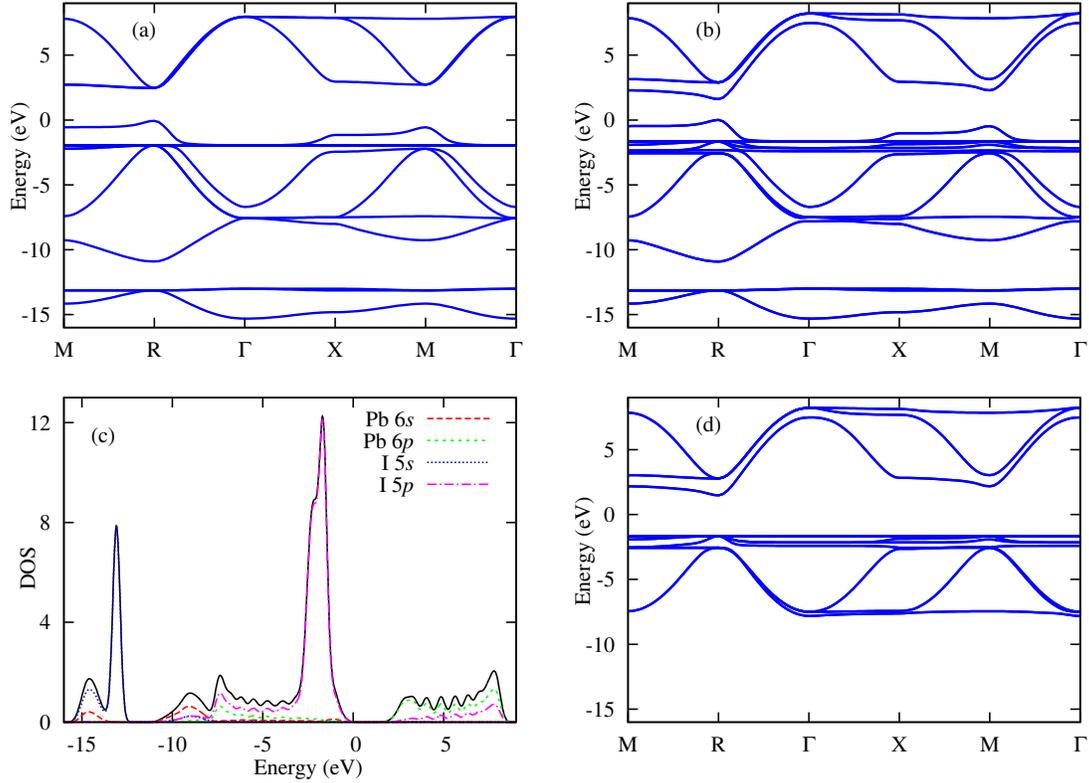}
\caption{Band structure of the tight-binding model without SOC (a) and including SOC (b). The quasi-momentum points marked on the x axis are $\Gamma=(0,0,0)$, X=(0.5,0,0), M=(0.5,0.5,0), and R=(0.5,0.5,0.5) [in units of $2\pi/a$, where $a$ is the lattice constant]. From (b) we calculate the density of states (DOS) per unit cell, which we plot in (c) (solid black line) along with the contributions of the different orbitals in the model: Pb 6$s$ (red dashed line), Pb 6$p$ (green short-dashed line), I 5$s$ (blue dotted line) and I 5$p$ (magenta dash-dotted line). In (d) we plot the band structure that we would obtain if we keep only the $p$ orbitals in the model, i.e.~remove from the Hamiltonian all matrix elements related to the $s$ orbitals. Note that all of the plots in this figure are obtained using calculations on a single unit cell and not our $8 \times 8 \times 8$ supercell. For the DOS we take 8000 quasi-momentum points distributed uniformly in the first Brillouin zone, and we replace each energy level by a Gaussian peak of width 0.2 eV, which is chosen to produce a smooth curve rather than a large number of narrow peaks.}
\label{Fig:BandStructure}
\end{figure}

We now present the results of the numerical calculations explained above. For reference and in order to have some insight about the system in the tight-binding model, we start with some general considerations. In Fig.~\ref{Fig:BandStructure}(b) we show the band structure in the absence of disorder, calculated using a single unit cell and a variable quasi-momentum. In Fig.~\ref{Fig:BandStructure}(a) we show the band structures that one would obtain if SOC is ignored. As can be seen from the figure the most obvious effect of including SOC is to create a splitting in the conduction band and reduce the band gap, compared to the case where SOC is ignored. In Fig.~\ref{Fig:BandStructure} we show the density of states (DOS) as a function of energy, along with the contributions from the different orbitals, for the case including SOC. The DOS shows that the $s$ orbitals contribute mostly to states at the bottom of the band structure in our model. To investigate whether these orbitals need to be kept in the model, we calculate the band structure for a model where the $s$ orbitals are ignored and we plot the results in Fig.~\ref{Fig:BandStructure}(d). One can see clearly that ignoring the $s$ orbitals strongly modifies the valence band, including the states that determine the band gap of the material. The $s$ orbitals therefore need to be kept in the model.

It is interesting to note that the band structure in Fig.~\ref{Fig:BandStructure}(b) does not exhibit a Rashba splitting, although evidence of such a splitting has been observed in recent experiments and it is believed to be related to SOC \cite{Niesner}. The reason is that the Rashba splitting requires both SOC and inversion symmetry breaking \cite{Kim,Mosconi}. Our model does not contain any inversion symmetry breaking. In the supplementary material we break the symmetry and show that the Rashba splitting appears in that case. The exact microscopic mechanism for inversion symmetry breaking in the materials under study is presently not well understood. Once this mechanism is established, the tight-binding model can be modified to represent it accurately. The Rashba splitting should not affect our results on the localization of electronic states presented below.

The energy levels in Fig.~\ref{Fig:BandStructure}(a-c) can be divided into three bands: a deep valence band (to which we shall refer as VB$-1$) that lies a few eV below the valence band, the valence band (VB) the conduction band (CB).

\subsection*{Disorder in on-site energies}

\begin{figure}[h]
\includegraphics[width=11.0cm]{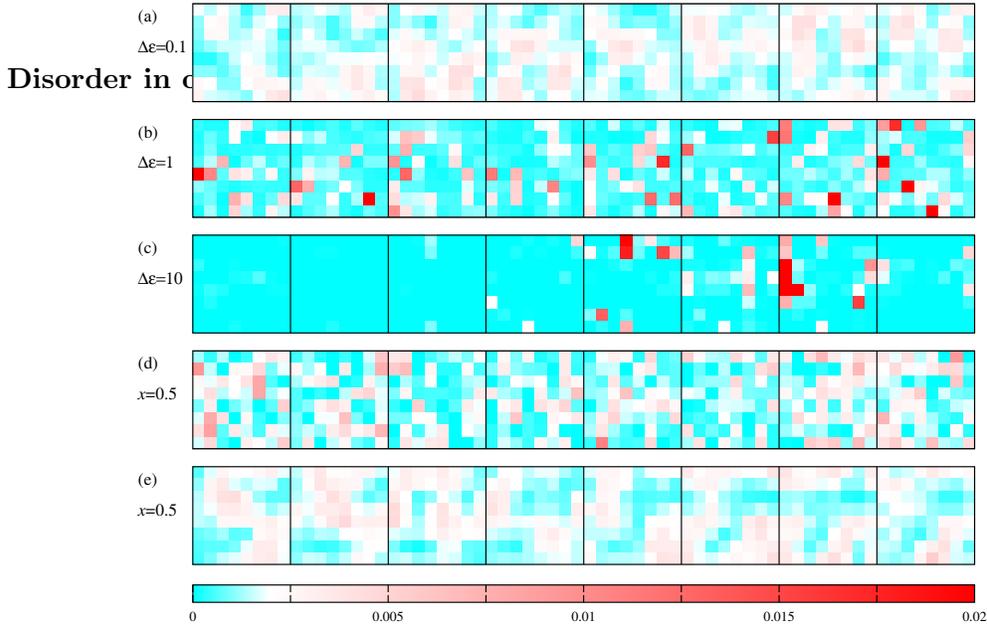}
\caption{Probability distribution for different disorder settings for the state with index $n=12,312$ which lies inside the valence band 1.66 eV below the VBM. Each panel has eight squares that correspond to the eight layers of unit cells in our $8 \times 8 \times 8$ supercell. Each one of the eight squares is divided into $8 \times 8$ small squares that correspond to the $8 \times 8$ unit cells in that layer. In other words, each panel has 512 small squares with different colours, representing the occupation probabilities of the 512 unit cells in our simulated supercell. A white square corresponds to a probability 1/512 (meaning that a uniform probability distribution would give an all-white figure), while blue corresponds to lower probabilities and red corresponds to higher probabilities (such that strong colours, especially with a mostly blue background and a few red squares, imply strongly localized states). Panels (a)-(c) correspond to on-site disorder with $\Delta\epsilon=$0.1 eV (a), 1 eV (b) and 10 eV (c). The tendency for the wavefunction to localize with increasing $\Delta\epsilon$ is clear from these panels. Panels (d) and (e) show the probability distribution for disorder originating from halide mixing. Here we assume disorder in the hopping strengths, and we take a 50-50 mixture of I and Br. In Panel (d) we show the results for the model where the Br orbital energies are different from those of I, while in Panel (e) we show the results for the model where the hopping strength between Pb and Br is different from that between Pb and I.}
\label{Fig:ProbabilityDistribution}
\end{figure}

We would like to investigate the localization properties of the energy eigenstates. First, to visualize the localization process, we plot in Fig.~\ref{Fig:ProbabilityDistribution} the probability distribution for a few different disorder settings. As the disorder strength increases, the states become increasingly localized and for any given state the occupation probability becomes concentrated in an increasingly small number of unit cells. In this figure and below, we shall go up to a disorder strength of 10 eV, which is higher than what we can expect in a realistic perovskite material, in order to show the behaviour in the limit where the electronic states becomes localized to a single or very few unit cells.

\begin{figure}[h]
\includegraphics[width=12.0cm]{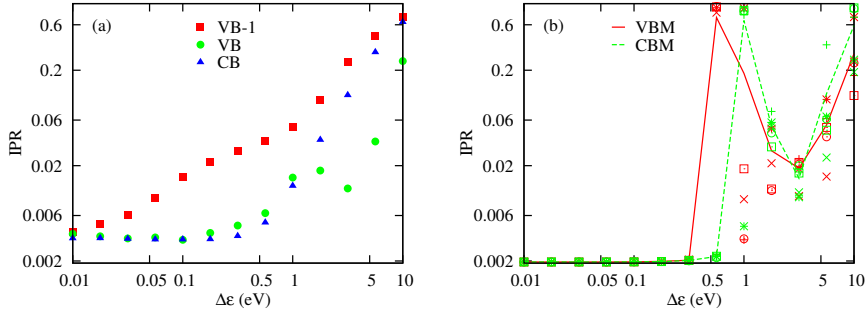}
\caption{The IPR as a function of the on-site disorder strength $\Delta\epsilon$. In Panel (a) we plot the mean value of the IPR for the three energy bands: VB$-1$ (red squares), VB (green circles) and CB (blue triangles). The IPR for all three bands increases gradually as $\Delta\epsilon$ approaches and exceeds 1 eV. We note here that although we are plotting the mean values over five instances, instance-to-instance fluctuations were small and almost indiscernible in the majority of cases. In Panel (b) we plot, for five instances, the IPR for the states at the VBM (red symbols) and CBM (green symbols). When we calculate the IPR for the VBM and CBM, we take the averages over the states that lie within 0.1 eV from the band edges. The solid and dashed lines are the respective mean values for the five instances. We can see that both lines increase and decrease again as $\Delta\epsilon$ increases. The counter-intuitive decrease occurs in the regime where there is no boundary between the VB and CB, and this delocalization tendency should therefore not be taken as reflecting the transport properties at the VBM and CBM. In (b) we see that the tendency towards localization is more sudden than when taking the mean values for entire energy bands. The localization of the states at the VBM and CBM occurs around $\Delta\epsilon\sim 0.2$-1 eV, with the VBM states becoming localized for somewhat smaller values of $\Delta\epsilon$.}
\label{Fig:IPRAsFunctionOfDelta}
\end{figure}

In Fig.~\ref{Fig:IPRAsFunctionOfDelta} we plot the IPR as a function of $\Delta\epsilon$. The IPR for both the VB and CB remains close to zero all the way up to $\Delta\epsilon\sim 0.2$ eV, after which it starts increasing and approaches unity for large values of $\Delta\epsilon$. If we take the mean values of the IPR over entire energy bands, the increase of the IPR is gradual. The VB$-1$ is the most sensitive band to disorder, and its IPR grows the fastest among the three bands. This feature can be understood by considering that this band is formed from $s$ orbitals, which have a hopping strength that is smaller than the hopping strength between $p$ orbitals by a factor of 3-4. The result that the electronic stats become localized when the disorder level reaches the range 0.2-1 eV can be understood intuitively by considering that each band in the band structure in Fig.~\ref{Fig:BandStructure}(b) has a width on the order of 1 eV. As a result, one can intuitively expect that localization will occur when the disorder strength becomes comparable to the energetic width of the bands.

The IPR for the states at the VBM and CBM remains very close to 1/512 up to $\Delta\epsilon\sim$ 0.3-0.5 eV [Fig.~\ref{Fig:IPRAsFunctionOfDelta})(b)]. The insensitivity of these states to small amounts of disorder can be understood based on the fact that the DOS is very small at the VBM and CBM [see Figs.~\ref{Fig:BandStructure}(c) and SI1(a)]. The disorder-induced localization can be understood as a hybridization between different basis states caused by the disorder when seen as a perturbation to the Hamiltonian in the zero-disorder limit. Energy levels that are far from other energy levels tend to hybridize less with other energy levels as a result of perturbations in the Hamiltonian. This insensitivity of the VBM and CBM states to disorder will be more clearly visible when we discuss halide mixing below. Once $\Delta\epsilon$ reaches 0.5 eV or higher, however, the increase in the IPR of the VBM and CBM state is quite abrupt, which might reflect the onset of a localization phase transition. The VBM states become localized for a somewhat smaller value of $\Delta\epsilon$ compared to the CBM states, which implies that hole transport should be more sensitive to on-site energy fluctuations than electron transport.

\subsection*{Halide mixing}

\begin{figure}[h]
\includegraphics[width=13.0cm]{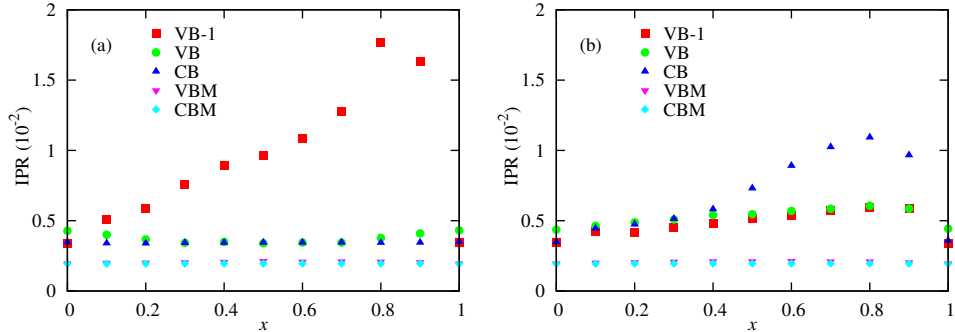}
\caption{The IPR as a function of Br concentration $x$. Panels (a) and (b) correspond, respectively, to the models where the orbital energies and hopping strengths are modified when an I atom is replaced by a Br atom. The different data sets correspond to the mean values of the IPR over the VB$-1$ (red squares), VB (green circles) and the CB (blue triangles), as well as the IPR values for for VBM (magenta inverted triangles) and CBM (cyan diamonds).}
\label{Fig:IPRAsfunctionOfHalideConcentrations}
\end{figure}

We now consider the case of disorder resulting from halide mixing, specifically the substitution of Br atoms for some of the I atoms in MAPbI$_3$ to give MAPbI$_{(3-x)}$Br$_x$. The IPR as a function of Br concentration $x$ is plotted in Fig.~\ref{Fig:IPRAsfunctionOfHalideConcentrations}. We distinguish between five different IPR values: the mean values over all states in the VB$-1$, VB and CB and the IPR values of the states at the VBM and CBM. The mean values over the occupied states can reflect the bond strengths in the material: delocalized states typically correspond to stronger bonding. In the case where the orbital energies change when substituting Br for I, the mean value of the IPR over the VB$-1$ states exhibits the largest variations as $x$ is varied, with the dependence being weak for small values of $x$ but with a large peak at $x=0.8$. In the model where the hopping strengths change when substituting Br for I, all three bands exhibit some increase in the IPR for mixed-halide materials with the most pronounced increases occurring at $x=0.8$. These results suggest that halide mixing with relatively small Br concentrations is easier than mixing with small I concentrations. The IPR values near the band edges reflect the transport properties of electrons and holes. We find that (in both disorder models) the IPR for states near the VBM increases by about 10\% in 50-50 mixed-halide materials as compared to pure I or pure Br materials, which suggests that the hole mobility will be almost unaffected by halide mixing. The IPR for states near the CBM is essentially independent of halide concentrations, suggesting that the disorder resulting from halide mixing has a lower effect on the electron mobility than it does on the hole mobility. This result makes sense, because the states near the CBM are composed mainly of Pb orbitals.

We note here that recent experiments suggest that doping with Cl might improve transport properties \cite{Colella,Wehrenfennig}. However, these results are widely believed to be related to defect formation during material synthesis. Such defects are not included in our model, and therefore we would not expect to capture these effects in our calculations.

\section*{Discussion}

In order to discuss the relevance of our results on on-site disorder to the materials in the family of MAPbI$_3$, we need to have an estimate for the parameter $\Delta\epsilon$ in these materials. Fluctuations related to the fluctuating orientation of the MA molecule were investigated in Ref.~\cite{Carignano2015} using molecular dynamics simulations, where the energetic barrier for the MA molecule rotation was found to be around 10-15 meV. Reference \cite{MadjetCharges} calculated the partial charges on the different atoms in the MA molecule in MAPbI$_3$. Using these charges and the relative positions of the atoms in the material, Coulomb's law gives that the rotation of the MA molecule would produce energy fluctuations on the order of 0.2-0.3 eV at the locations of the neighbouring atoms. If we include the fact that the dielectric constant is on the order of 10, the charges on the MA molecule will be screened by a similar factor, and we again obtain the similar estimate $\Delta\epsilon\sim 0.02$ eV. Figure \ref{Fig:IPRAsFunctionOfDelta} shows that the states start to become localized around $\Delta\epsilon=0.2$ eV, which is about one order of magnitude larger than the value of $\Delta\epsilon$ that results from the MA orientation fluctuations. We note here that Ref.~\cite{Ma} estimated potential fluctuations to be in the range 0.2-0.5 eV. These numbers are consistent with the electronic state localization found there.

Our discussion above has concentrated on the IPR, which quantifies the localization of the energy eigenstates of the system. This localization should be reflected in the transport properties. In particular, the mobility is defined as
\begin{equation}
\mu = \frac{v_d}{E_{\rm app}},
\end{equation}
where $v_d$ is the drift velocity obtained upon applying an electric field of strength $E_{\rm app}$. For a given (incoherent) scattering rate, the drift velocity can be expected to scale as the spatial size of the relevant wave functions. The argument here is that after a given scattering event the charge carrier moves ballistically until the next scattering event, and when the electronic states have a certain characteristic spatial size the electrons will remain within the same length scale between scattering events. As a result, one can expect that the mobility is proportional to IPR$^{-1/3}$. These relations agree with the intuitive expectation that localization of the energy eigenstates, which leads to increased values of the IPR, leads to a reduction in the mobility.

To conclude, motivated by the emerging lead halide perovskite materials and their remarkable transport properties, we have analyzed the localization properties of the energy eigenstates of a tight-binding model that contains disorder stemming from random MA molecule orientations or from halide mixing. The IPR increases as the on-site disorder strength is increased, indicating a degradation of transport properties. However, our results suggest that the disorder strength in the lead halide perovskites puts them in the regime of delocalized states, where the mobility is not significantly reduced because of the disorder. We also find that the disorder originating from halide mixing has a minimal effect on the IPR near the band edges and therefore on the mobility. However, we do observe more dramatic changes deeper in the VB, which could indicate weaker bonds for the mixed-halide perovskites, which in turn is consistent with the phase separation that has been observed experimentally. Combined with other favourable material properties, such as the absence of deep trap states, our results help explain the good transport properties of these materials.


We would like to thank M. Carignano for useful discussions. This work was made possible by NPRP grant \# 8-086-1-017 from the Qatar National Research Fund (a member of Qatar Foundation). The findings achieved herein are solely the responsibility of the authors.

Author contributions: S.A. formulated the problem, performed the numerical calculations and played the leading role in writing the manuscript. M.E.M performed the calculations for the parameter estimation and contributed to the model formulation, interpretation of the results and writing of the manuscript. O.V., S.H. and E.H.S. contributed to the interpretation of the results and writing of the manuscript.

Competing financial interests: The authors declare no competing financial interests.

Correspondence to: Sahel Ashhab, sashhab@hbku.edu.qa

\begin{center}
\textbf{\large Supplementary Information}
\end{center}

Here we present several additional results about the lead-halide perovskites obtained from the tight-binding model. Some of these results are closely related to our investigation of localization effects in disordered materials, while others are more general results pertaining to the tight-binding model of the lead-halide perovskites.

\section*{Rashba splitting}

\begin{figure}[h]
\includegraphics[width=8.0cm]{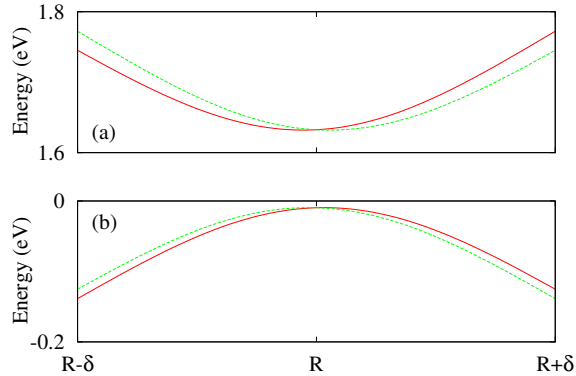}
\caption{Band structure of the tight-binding model with SOC and inversion symmetry breaking. We show only the energy levels at the CBM (a) and VBM (b), which are located at the R point. The range of the x axis is given by the parameter $\delta=(0.05,0,0)$. The single lines of the VBM and CBM in Fig.~2 of the main text are now split into two lines each.}
\label{Fig:BandStructureWithBrokenInversionSymmetry}
\end{figure}

As mentioned in the main text, the band structure in Fig.~2 does not exhibit a Rashba splitting at the CBM and VBM because the Hamiltonian that are using in this work does not break inversion symmetry. We now show that if this symmetry is broken, a Rashba splitting is obtained. We modify the system parameters such that they are all kept unchanged except for the hopping strength between the Pb $p$ orbitals and the I $p$ orbitals when the I atom is below the Pb atom in the z direction, a change that makes the system not mirror symmetric about the xy plane anymore and hence breaks inversion symmetry in the z direction. The hopping between aligned $p$ orbitals (i.e.~$\sigma$ configuration) is modified to 0.9 of the original value, and the hopping between parallel $p$ orbitals (i.e.~$\pi$ configuration) is modified to 1.5 of the original value. We needed to make two parameter modifications in order to be able to independently adjust the Rashba splitting in the CB and VB: the former modification splits the bands at the VBM and CBM by equal amounts, while the latter modification contributes to the splitting at the VBM and causes a much smaller change at the CBM. The energy level spectrum around the R point, where the VBM and CBM are located, is plotted in Fig.~\ref{Fig:BandStructureWithBrokenInversionSymmetry}.

\section*{Energy levels and IPR as functions of level index}

\begin{figure}[h]
\includegraphics[width=15.0cm]{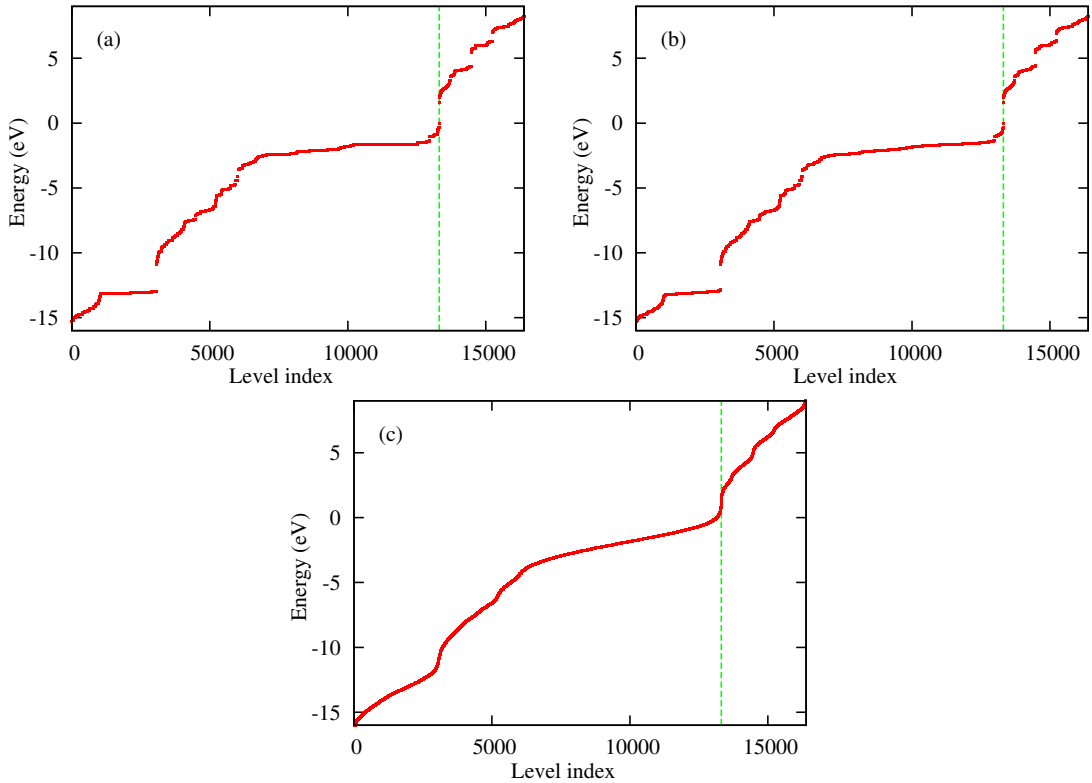}
\caption{Energy levels for an $8 \times 8 \times 8$ supercell at zero quasi-momentum as a function of energy level index when including disorder in the on-site energies. The disorder values used are $\Delta\epsilon=0$ (a), 0.1 eV (b) and 1 eV (c). The energy level index ranges from 1 to 16,384, which is the number of electronic states in one supercell. The green vertical line marks the location of the VBM ($n$=13,312) and CBM ($n$=13,313). Panels (a) and (b) have another gap between the DVB (up to $n$=3,072) and the VB (starting from $n$=3,073). The figures plotted here are produced from single instances of the disordered potential, but we find that all the disorder instances that we have generated (i.e.~five instances for each value of $\Delta\epsilon$) give results that look almost identical to those shown here.}
\label{Fig:EnergyLevelsWithOnSiteDisorder}
\end{figure}

In Fig.~\ref{Fig:EnergyLevelsWithOnSiteDisorder}(a) we plot the energy levels of a $8 \times 8 \times 8$ supercell as a function of level index, in the absence of disorder and taking the quasimomentum to be zero. In other words, the energy levels obtained from diagonalizing the zero-quasimomentum, periodic-boundary-condition Hamiltonian are arranged in increasing order and plotted. Taking a large supercell and considering only zero quasi-momentum is in certain ways equivalent to taking a single unit cell but considering a large number of uniformly distributed quasi-momentum values. Indeed one can see that the DOS in Fig.~2(c) in the main text reflects the energy level distribution in Fig.~\ref{Fig:EnergyLevelsWithOnSiteDisorder}(a). 

\begin{figure}[h]
\includegraphics[width=8.0cm]{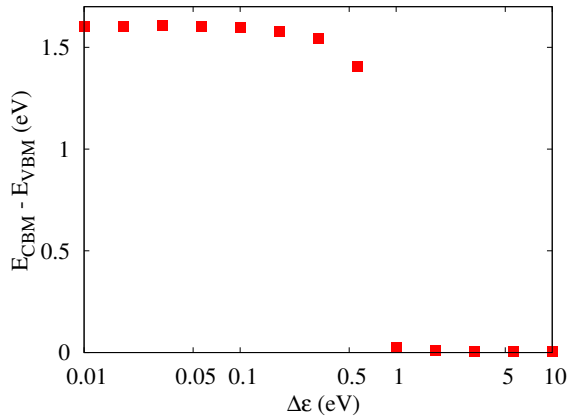}
\caption{The energy difference between level number 13,313 and level number 13,312 as a function of $\Delta\epsilon$. For small values of $\Delta\epsilon$, these two energy levels correspond to the CBM and VBM, respectively. The plotted quantity is therefore the band gap of the material. As can be seen in Fig.~\ref{Fig:EnergyLevelsWithOnSiteDisorder}(c), for $\Delta\epsilon\gtrsim 1$ the band structure is lost. For such large values of $\Delta\epsilon$, the quantity plotted here loses its meaning as the band gap.}
\label{Fig:BandGapAsFunctionOfDelta}
\end{figure}

We now include disorder in the on-site energies. In Fig.~\ref{Fig:EnergyLevelsWithOnSiteDisorder} we plot the energy levels as functions of level index for three different values of the disorder strength $\Delta\epsilon$. In Fig.~\ref{Fig:BandGapAsFunctionOfDelta} we plot the energy difference between the two states at the CBM ($n$=13,313) and VBM ($n$=13,312). For small values of $\Delta\epsilon$, up to around 0.2 eV, the on-site energy disorder has little effect on the energy levels. As $\Delta\epsilon$ is increased further and reaches the range 0.5-1 eV, the energy levels are modified drastically by the disorder and the energy gap disappears. If we increase $\Delta\epsilon$ further to 5 eV or higher, the energy levels form a distribution whose statistical properties reflect those of the Gaussian on-site energy fluctuations, hence losing all features that are specific to the underlying perovskite lattice. Note that the shrinking and eventual disappearance of the gap in the energy levels does not mean that the material will absorb photons of smaller frequencies and eventually become a conductor, because when we reach the point where the energy levels form a continuum the corresponding wavefunctions become separated spatially and one would not observe direct transitions between them.

\begin{figure}[h]
\includegraphics[width=14.0cm]{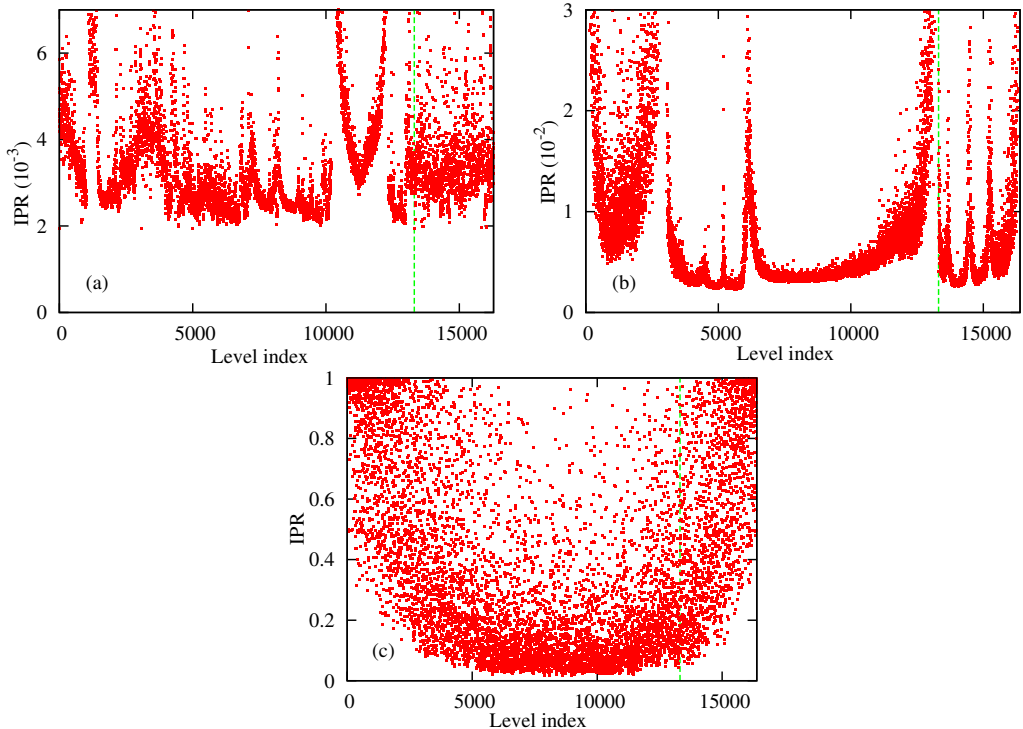}
\caption{Inverse participation ratio (IPR) for the energy eigenstates of an $8 \times 8 \times 8$ supercell as a function of energy level index in the presence of on-site energy fluctuations. The different panels correspond to different values of the disorder strength: $\Delta\epsilon=0.01$ eV (a), 1 eV (b) and 10 eV (c). The green vertical line marks the location of the VBM and CBM. The figures are produced from single instances of the disordered potential, but we find that the overall features are the same for all five disorder instances that we have generated for each value of $\Delta\epsilon$.}
\label{Fig:IPRWithOnSiteDisorder}
\end{figure}

\begin{figure}[h]
\includegraphics[width=14.0cm]{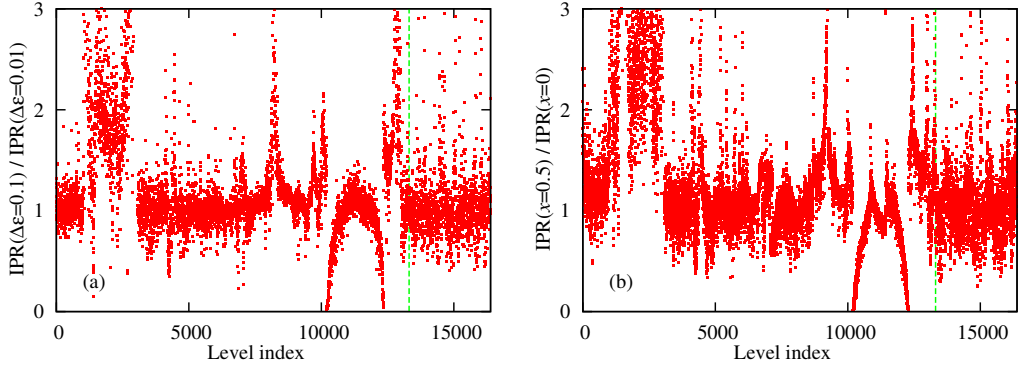}
\caption{The ratios IPR($\Delta\epsilon=0.1$ eV)/IPR($\Delta\epsilon=0.01$ eV) (Panel a) and IPR($x=0.5$)/IPR($x=0$) (Panel b). In (b) we use the model where Br orbital energies are different from those of I orbitals. The green vertical line marks the location of the VBM and CBM. This figures shows that some states become more localized while others become less localized when increasing the amount of disorder, keeping in mind that for very strong disorder all states become localized.}
\label{Fig:IPRRatios}
\end{figure}

In Fig.~\ref{Fig:IPRWithOnSiteDisorder} we plot the IPR as a function of level index for three values of $\Delta\epsilon$. For small values of $\Delta\epsilon$, all the energy eigenstates are generally delocalized, although there are large variations in their IPR values. As $\Delta\epsilon$ is increased, the IPR generally increases and the states become more localized. Interestingly, this is not true for all the states, and some states become more delocalized. This point is illustrated in Fig.~\ref{Fig:IPRRatios}(a), where we plot the ratio between the IPR at $\Delta\epsilon=0.1$ eV and the IPR at $\Delta\epsilon=0.01$ eV. The figure shows that the majority of the electronic states become more localized as $\Delta\epsilon$ is increased. However, some states, particularly near two energies inside the valence band, become less localized with increasing disorder. Another point to note in this figure is that the states around the VBM and CBM do not change drastically when the disorder is increased to 0.1 eV. Naturally, for very strong disorder all states become localized, as can be seen in Fig.~\ref{Fig:IPRWithOnSiteDisorder}(c).

We do not show the detailed plots of the IPR as a function of level index $n$, because it looks similar to that shown in Fig.~\ref{Fig:IPRWithOnSiteDisorder}(a). Similarly, the ratio between the IPR at $x=0.5$ and the IPR at $x=0$ as a function of energy level index is plotted in Fig.~\ref{Fig:IPRRatios}(b), and it exhibits similar features to those obtained in the case of on-site disorder.

\section*{Mixed-halide material band gap}

\begin{figure}[h]
\includegraphics[width=8.0cm]{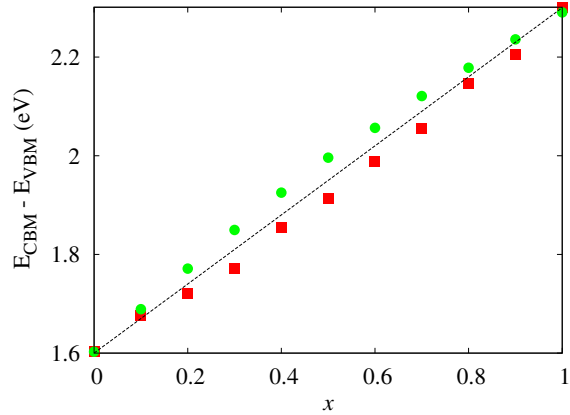}
\caption{Band gap as function of Br concentration $x$ (where Br is treated as a dopant replacing I in MAPbI$_3$). The red squares and green circles correspond, respectively, to the models where the orbital energies and hopping strengths are modified when an I atom is replaced by a Br atom. The dashed line is a straight line the goes from 1.6 eV at $x=0$ to 2.3 eV at $x=1$. In both cases, the band gap increases almost linearly from 1.6 eV for MAPbI$_3$ to 2.3 eV for MAPbBr$_3$.}
\label{Fig:MixedHalideBandGap}
\end{figure}

In Fig.~\ref{Fig:MixedHalideBandGap} we plot the band gap of the mixed-halide perovskite MAPbI$_{3(1-x)}$Br$_{3x}$ as a function of Br concentration $x$. The band gap increases almost linearly from 1.6 eV for MAPbI$_3$ to 2.3 eV for MAPbBr$_3$.


\begin{thebibliography}{99}

\bibitem{Kojima} A. Kojima, K. Teshima, Y. Shirai, and T. Miyasaka, Organometal Halide Perovskites as Visible-Light Sensitizers for Photovoltaic Cells, J. Am. Chem. Soc. {\bf 131}, 6050 (2009).

\bibitem{PerovskiteReview} M. A. Green, A. Ho-Baillie, and H. J. Snaith, The emergence of perovskite solar cells, Nature Photon. {\bf 8}, 506 (2014).

\bibitem{PerovskiteTransportReview} T. M. Brenner, D. A. Egger, L. Kronik, G. Hodes, and D. Cahen, Hybrid organic-inorganic perovskites: low-cost semiconductors with intriguing charge-transport properties, Nature Rev. Mat. {\bf 1}, 15007 (2016).

\bibitem{Edri} E. Edri, S. Kirmayer, D. Cahen, and G. Hodes, High Open-Circuit Voltage Solar Cells Based on Organic-Inorganic
Lead Bromide Perovskite, J. Phys. Chem. Lett. {\bf 4}, 897 (2013).

\bibitem{McMeekin} D P. McMeekin {\it et al.}, A mixed-cation lead mixed-halide perovskite absorber for tandem solar cells, Science {\bf 351}, 6269 (2016).

\bibitem{Poglitsch} A. Poglitsch and D. Weber, Dynamic disorder in methylammoniumtrihalogenoplumbates (II) observed
by millimeter-wave spectroscopy, J. Chern. Phys. {\bf 87}, 6373 (1987).

\bibitem{Mashiyama} H. Mashiyama {\it et al.}, Disordered configuration of methylammonium in CH$_3$NH$_3$PbBr$_3$ determined by single crystal neutron diffractometry, Ferroelectrics {\bf 348}, 182 (2007).

\bibitem{Frost} J. M. Frost, K. T. Butler, and A. Walsh, Molecular ferroelectric contributions to anomalous hysteresis in hybrid perovskite solar cells APL Mater. {\bf 2}, 081506 (2014).

\bibitem{Carignano2015}  M. A. Carignano, A. Kachmar, and J. Hutter,  Thermal Effects on CH$_3$NH$_3$PbI$_3$ Perovskite from Ab initio Molecular Dynamics Simulations, J. Phys. Chem. C {\bf 119}, 8991 (2015).

\bibitem{Leguy} A. M. A. Leguy {\it et al.}, The dynamics of methylammonium ions in hybrid organic-inorganic perovskite solar cells, Nature Commun. {\bf 6}, 7124 (2015).

\bibitem{Bakulin} A. A. Bakulin {\it et al.}, Real-Time Observation of Organic Cation Reorientation in Methylammonium Lead Iodide Perovskites, J. Phys. Chem. Lett. {\bf 6}, 3663 (2015).

\bibitem{Ma} Jie Ma and Lin-Wang Wang, Nanoscale Charge Localization Induced by Random Orientations of Organic Molecules in Hybrid Perovskite CH$_3$NH$_3$PbI$_3$, Nano Lett. {\bf 15}, 248 (2015).

\bibitem{Laovorkiat} C. La-o-vorakiat {\it et al.}, Elucidating the role of disorder and free-carrier recombination kinetics in CH3NH3PbI3 perovskite films, Nature Commun. {\bf 6}, 7903 (2015).

\bibitem{Yaffe} O. Yaffe {\it et al.}, The nature of dynamic disorder in lead halide perovskite crystals, arXiv:1604:08107.

\bibitem{Buin} A. Buin {\it et al.}, Materials Processing Routes to Trap-Free Halide Perovskites, Nano Lett. {\bf 14}, 6281 (2014).

\bibitem{Anderson} P. W. Anderson, Absence of Diffusion in Certain Random Lattices, Phys. Rev. {\bf 109}, 1492 (1958).

\bibitem{Berry} J. Berry {\it et al.}, Hybrid Organic-Inorganic Perovskites (HOIPs): Opportunities and Challenges, Adv. Mater. {\bf 27}, 5102 (2015).

\bibitem{Brandt} R. E. Brandt, V. Stevanovic, D. S. Ginley, and T. Buonassisi, Identifying defect-tolerant semiconductors with high minority carrier lifetimes: Beyond hybrid lead halide perovskites, MRS Communications {\bf 5}, 265 (2015).

\bibitem{Goringe} C. M. Goringe, D. R. Bowler, and E. Hernandez, Tight-binding modelling of materials, Rep. Prog. Phys. {\bf 60}, 1447 (1997).

\bibitem{Agapito} L. A. Agapito, S. Ismail-Beigi, S. Curtarolo, M. Fornari, and M. Buongiorno Nardelli, Accurate tight-binding Hamiltonian matrices from ab-initio calculations: Minimal basis sets, Phys. Rev. B {\bf 93}, 035104 (2016).

\bibitem{BoyerRichard} S. Boyer-Richard, C. Katan, B. Traor\'e, R. Scholz, J.-M. Jancu, and J. Even, Symmetry-Based Tight Binding Modeling of Halide Perovskite Semiconductors, J. Phys. Chem. Lett. {\bf 7}, 3833 (2016).

\bibitem{Kim} M. Kim, J. Im, A. J. Freeman, J. Ihm, and H. Jin, Switchable S=1/2 and J=1/2 Rashba bands in ferroelectric halide perovskites, Proc. Natl. Acad. Sci. {\bf 111}, 6900 (2014).

\bibitem{MadjetThermalization} M. E. Madjet {\it et al.}, Enhancing the Carrier Thermalization Time in Organometallic Perovskites by Halide Mixing, Phys. Chem. Chem. Phys. {\bf 18}, 5219 (2016).

\bibitem{Mattoni} A. Mattoni, A. Filippetti, M. I. Saba, and P. Delugas, Methylammonium rotational dynamics in lead halide perovskite by classical molecular dynamics: the role of temperature, J. Phys. Chem. C 119, 17421–17428 (2015).

\bibitem{Lahnsteiner} J. Lahnsteiner, G. Kresse, A. Kumar, D. D. Sarma, C. Franchini, M. Bokdam, Room-temperature dynamic correlation between methylammonium molecules in lead-iodine based perovskites: An ab initio molecular dynamics perspective, Phys. Rev. B {\bf 94}, 214114, (2016).

\bibitem{Meloni} S. Meloni {\it et al.}, Ionic polarization-induced current–voltage hysteresis in CH$_3$NH$_3$PbX$_3$ perovskite solar cells, Nat. Commun. {\bf 7}, 10334 (2016).

\bibitem{Carignano2017} M. Carignano {\it et al.} (unpublished).

\bibitem{Bell} R. J. Bell and P. Dean, Atomic vibrations in vitreous silica, Discuss. Faraday Soc. {\bf 50}, 55 (1970).

\bibitem{Visscher} W. M. Visscher, Localization of electron wave functions in disordered systems, J. Non-Cryst. Sol. {\bf 8-10}, 477 (1972).

\bibitem{Ashhab} S. Ashhab, Quantum state transfer in a disordered one-dimensional lattice, Phys. Rev. A {\bf 92}, 062305 (2015).

\bibitem{Magnetta} B. Magnetta, G. Ordonez, S. Garmon, Impurity-directed transport within a finite disordered lattice, arXiv:1511.08758.

\bibitem{Niesner} D. Niesner, M. Wilhelm, I. Levchuk, A. Osvet, S. Shrestha, M. Batentschuk, C. Brabec, and T. Fauster, Giant Rashba Splitting in CH$_3$NH$_3$PbBr$_3$ Organic-Inorganic Perovskite, Phys. Rev. Lett. {\bf 117}, 126401 (2016).

\bibitem{Mosconi} E. Mosconi, T. Etienne, and F. De Angelis, Rashba Band Splitting in Organohalide Lead Perovskites: Bulk and Surface Effects, J. Phys. Chem. Lett. {\bf 8}, 2247 (2017).

\bibitem{Colella} S. Colella {\it et al.}, MAPbI$_{\rm 3-x}$Cl$_{\rm x}$ Mixed Halide Perovskite for Hybrid Solar Cells: The Role of Chloride as Dopant on the Transport and Structural Properties, Chem. Mater. {\bf 25}, 4613 (2013).

\bibitem{Wehrenfennig} C. Wehrenfennig {\it et al.}, High Charge Carrier Mobilities and Lifetimes in Organolead Trihalide Perovskites. Adv. Mater. {\bf 26}, 1584 (2014).

\bibitem{MadjetCharges} M. E. Madjet {\it et al.}, Atomic partial charges on CH3NH3PbI3 from first-principles electronic structure calculations, J. Appl. Phys. {\bf 119}, 165501 (2016).

\end{thebibliography}
\end{document}